\documentclass[a4paper,11pt]{article}
\usepackage{jinstpub} 

\title{Evaluation of the Front-End FERS 5202 Readout System for Muon Radiography Applications}

\author[a,b,c,1]{{R.M.I.D Gamage},\note{Corresponding author.}}
\author[a,b]{F. Ambrosino,}
\author[a,b]{L. Cimmino,}
\author[a,b]{G. Nyitrai,}
\author[a,b]{G. Saracino.}
\affiliation[a]{{University of Naples Federico II}\\Via Cintia, Napoli, Italy}
\affiliation[b]{Istituto Nazionale di Fisica Nucleare, sezione di Napoli,\\Via Cintia, Napoli, Italy}
\affiliation[c]{University of Ruhuna,\\Matara, Sri Lanka}

\emailAdd{darshanaishan@gmail.com}

\abstract{
This work presents a comprehensive characterization of the FERS 5202 front-end readout unit when processing signals from Silicon Photo-multipliers (SiPMs). The readout system's performance is characterized in terms of its charge resolution, dynamic range, and noise performance at the single photoelectron level, which is critical for applications requiring detection of low-light signals such as medical imaging, high-energy physics experiments and photon counting applications. The evaluation encompassed both fundamental performance metrics and practical implementation scenarios using Hamamatsu MPPC devices. This work aims to provide valuable reference data not only for our intended muon radiography application but also for the broader scientific community employing SiPMs in diverse experimental contexts.
}

\begin{document}
\date{}
\maketitle
\flushbottom

\section{Introduction}
\label{sec:intro}

Silicon Photo-multipliers (SiPMs) have emerged as critical photo detectors in numerous high-energy physics~\cite{Ahmed2025}, nuclear physics, and medical imaging applications~\cite{Park2022} due to their compact size, insensitivity to magnetic fields, low operating voltage, and excellent single-photon detection capabilities. However, fully exploiting these advantages requires sophisticated front-end electronics capable of addressing the unique challenges SiPMs present, including high dark count rates, temperature sensitivity, and the need for precise timing and charge measurements across a wide dynamic range.

\begin{figure}[htbp]
\centering
\includegraphics[width=.4\textwidth]{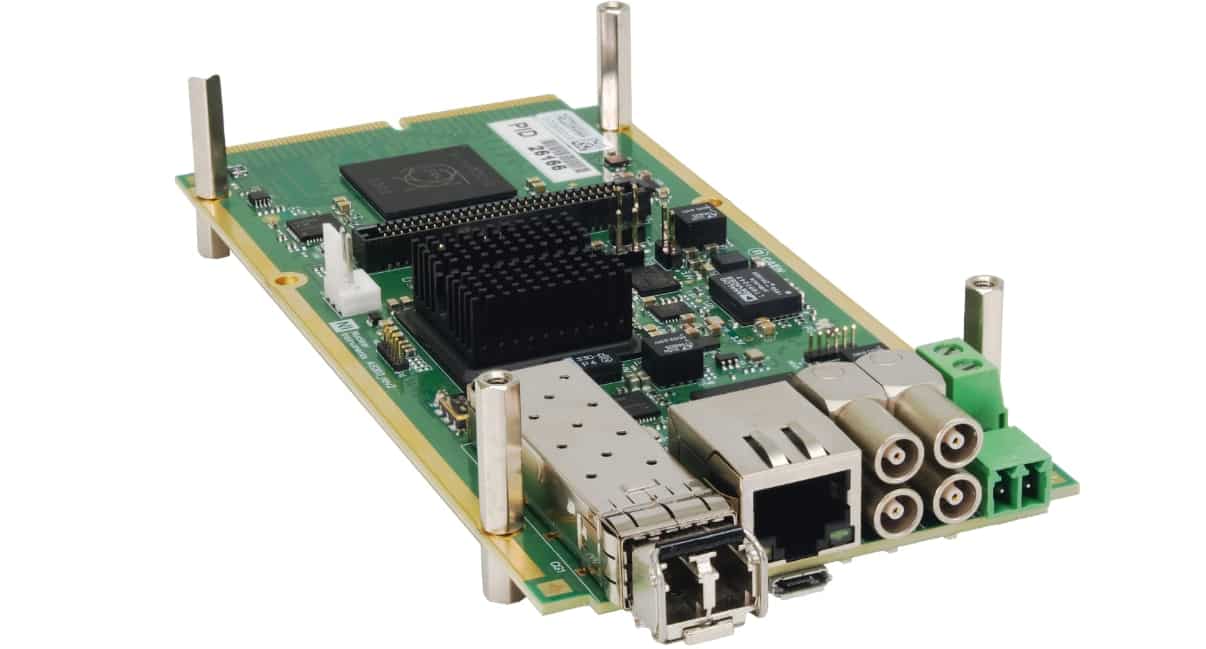}
\qquad
\includegraphics[width=.4\textwidth]{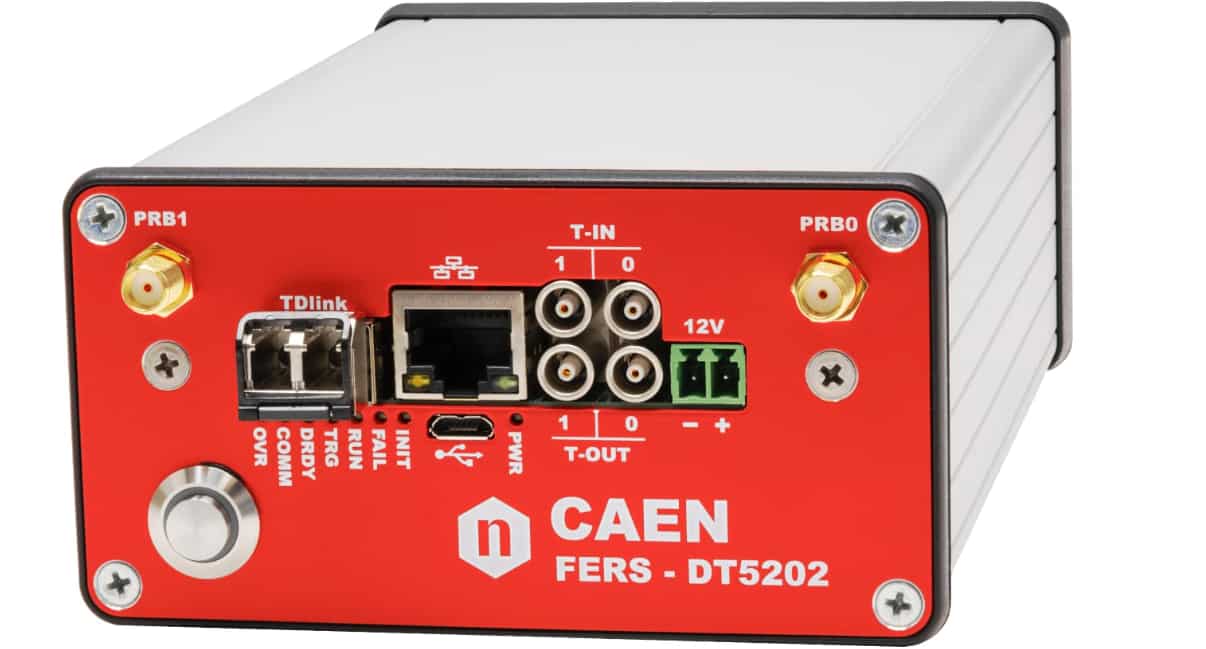}
\caption{FERS 5200, A5202 (left) and DT5202 desktop version (right) units.  \label{fig:i}}
\end{figure}

The A5202 unit, made by CAEN~\cite{CAEN2025}, is based on the 64 channels CITIROC~\cite{CAEN_a5202} front-end chip (Figure~\ref{fig:i}) and  represents a state-of-the-art solution for SiPM signal processing, combining high channel density with advanced signal processing capabilities. While several commercial readout solutions exist for SiPM applications~\cite{petsys_TOF_FEBD,aptech_photoniq_DAQ}, comprehensive characterization studies of these systems under realistic experimental conditions remain limited in the literature, particularly for specific applications.

This work presents a detailed characterization of the FERS 5200 readout electronics when processing signals from SiPMs in conjunction with fast plastic scintillator trigger for cosmic muons. Such a configuration is particularly relevant for applications requiring precise coincidence timing or where background discrimination is crucial. Our specific interest lies in the potential application of this system for muon radiography, where detector compactness, power efficiency, and reliable operation in challenging environmental conditions are essential requirements.

By thoroughly characterizing this electronic readout system, this work aims to provide valuable reference data not only for our intended muon radiography application but also for the broader scientific community employing SiPMs in diverse experimental contexts.

\section{ FERS-5200 Readout System: Technical Overview}
 
 FERS-5200 is a Front-End Readout System developed by CAEN SpA~\cite{CAEN2025} for multi-detector arrays. Its distributed, compact, and scalable architecture makes it particularly interesting for muon radiography applications, especially in challenging environments like bore hole detectors~\cite{Cimmino_2022} where space limitations and system durability are critical factors.

Each FERS-5200 Front-End unit is a compact card (approximately 15 × 6 cm) that integrates the complete signal processing chain including analog signal processing using Application-Specific Integrated Circuits (ASIC), analog-to-digital conversion, digital signal processing via FPGA, local data storage, synchronization capabilities, and communication interfaces~\cite{venaruzzo2020fers5200}. This architecture offers significant advantages for field-deployed muon radiography systems. Unlike conventional approaches that separate detectors from rack-mounted electronics with long analog signal cables, the FERS-5200 allows to position the entire signal processing chain adjacent to the detector. This design minimizes noise pickup and signal attenuation—critical for maintaining signal integrity in electrically noisy environments and across the extended cable runs typical in deep geological investigations.

\begin{figure}[htbp]
    \centering
    \includegraphics[width=.6\textwidth]{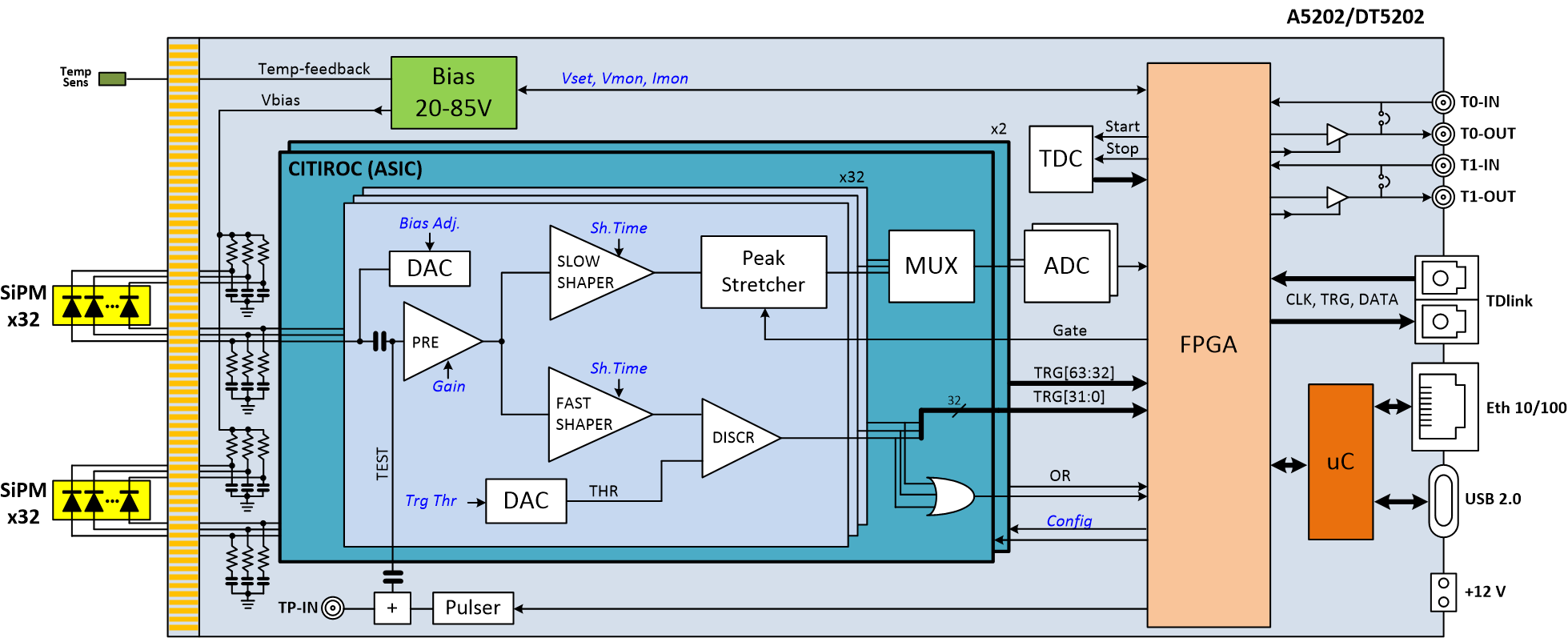}
    \caption{ Block diagram of the FERS 5202 Unit.~\cite{CAEN_a5202}}  \label{fig:block_fers5200}
\end{figure}

The system has very sophisticated networking capabilities and designed for detectors with a large number of channels that can be triggered externally as in collider experiments. Many units connect via daisy-chain through a high-speed optical TDlink operating at 6.25 Gbit/s, with up to 16 FERS units possible in a single chain—providing ample capacity for implementation. A single Concentrator Board (DT5215)~\cite{DT5215_manual} can manage the simultaneous readouts through a single TDlink, resulting in all the readout channels with precise synchronization for our experimental setup. The TDlink simultaneously provides data readout with bandwidth exceeding 100 MB/s shared between linked units, distribution of a reference clock for precise timing synchronization across all boards, and broadcasting of commands (triggers, time resets, etc.) to ensure coordinated data acquisition~\cite{CAEN_a5202}.

\begin{figure}
    \centering
    \includegraphics[width=0.6\linewidth]{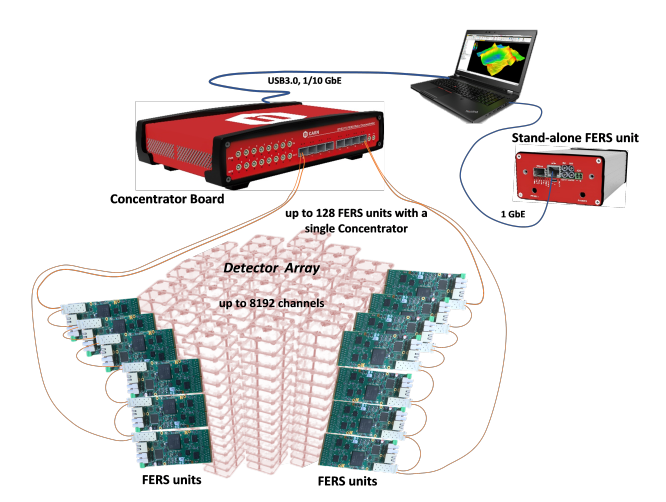}
    \caption{FERS-5200 Tree Network: Up to 16 FERS units can be connected in a daisy-chain configuration using the Optical Link (TDlink). A single Concentrator Board can manage data from up to 8 TDlink connections, supporting a maximum of 128 FERS units in total.~\cite{venaruzzo2020fers5200}
    \label{fig:daisy_chain}}
\end{figure}

The A5202 unit, specifically designed for SiPM array readout, provides key features beneficial for muon radiography applications through its 64 input channels implemented via two CITIROC 1A ASICs~\cite{weeroc_citiroc1a}. It features a programmable high voltage power supply (20-85 V, 10 mA) for SiPM bias with temperature for gain stability—both critical for maintaining consistent detector response in variable field conditions during long-term deployments. Channel-by-channel fine adjustment of high-voltage enables detector uniformity calibration, while high-resolution timing capabilities (<100 ps from CITIROC, 50 ps LSB from additional TDC) prove crucial for accurate muon trajectory reconstruction. The edge card connector (HSEC8-170) interfaces efficiently with SiPM arrays, minimizing connection points in space-constrained configurations. 

\begin{figure}
    \centering
    \includegraphics[width=0.8\linewidth]{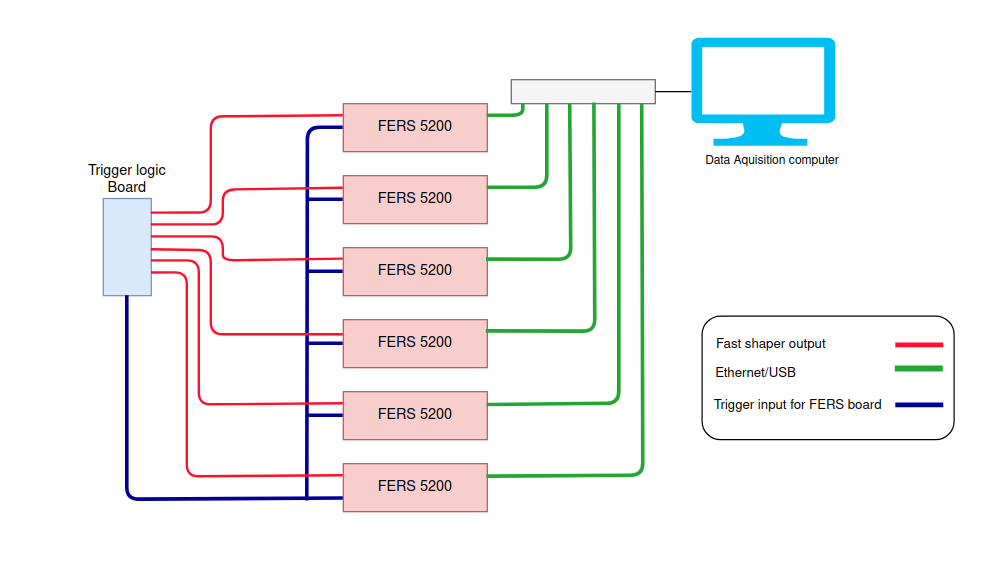}
    \caption{Sketch of the external trigger system using the fast shaper}
    \label{fig:ex_trig}
\end{figure}

For muon radiography detectors, the trigger is typically provided by the passage of the muon itself into the detector as illustrated in figure~\ref{fig:ex_trig}. We tested this possibility with good results~\cite{Bottiglieri2023}. For this study anyway, we used an external trigger produced by plastic scintillator bars coupled with PMT (see figures~\ref{fig:dark_box} and ~\ref{fig:setup_darkbox})  The fast shaper output is fed to an external trigger board, which generates trigger logic according to our requirements and feeds it back to the FERS 5200 board through the T0/T1 LEMO input for data acquisition. The FERS boards are connected to the data acquisition computer via USB 2.0 or Ethernet, as illustrated in the figure~\ref{fig:ex_trig}.

\section{Trigger Acceptance Frequency Characterization}

\begin{figure}
    \centering
    \includegraphics[width=0.4\linewidth]{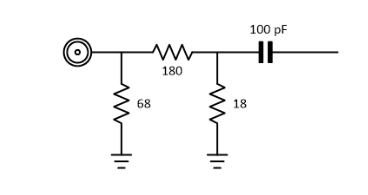}
    \caption{Charge injector circuit for the FERS A5202 unit.}
    \label{fig:Charge_injector_circuit}
\end{figure}

The trigger acceptance frequency of the FERS-A5202 system was evaluated under controlled laboratory conditions to assess its performance in processing signals from SiPMs under varying input rates. A Tektronix AFG 3252 function generator, configured in pulse mode, generated synchronized trigger and charge signals to emulate detector responses. The charge injection circuit (Figure~\ref{fig:Charge_injector_circuit}) shape produces pulses with 5 ns rise/fall times. The charge signal amplitude was set to 300 mV , while the trigger signal operated at 1.5 V logic levels (Table ~\ref{tab:set values}).

\begin{table}[ht]
\centering
\caption{Trigger and charge signal characteristics }
\begin{tabular}{|c|c|c|}
\hline
  Parameter & Trigger Signal & Charge signal \\ \hline
\hline
Frequency & 100 Hz & 100 Hz \\ \hline
 Delay & 30 ns  & 0 ns \\ \hline
High & 1.5 V & 300 mV \\ \hline
Low & 0 V & 0 mV \\ \hline
width & 100 ns & 100 ns \\ \hline
Leading edge & 5 ns & 5 ns \\ \hline
trailing edge & 5 ns & 5 ns\\ \hline
\end{tabular}

\label{tab:set values}
\end{table}

Trigger acceptance tests were conducted under two distinct configurations: (1) with only one channel active, and (2) with all 64 channels active. In both cases, input trigger frequencies ranged from 100 Hz to 30 kHz, with data acquisition performed over 120-second intervals to ensure statistical robustness. External triggering was employed to simulate the external scintillator trigger.

The trigger acceptance rate was defined as the ratio of successfully registered triggers to the total number of injected pulses. The system operated in spectroscopic mode i.e the charge amplitude is acquired and digitized, with channel activation managed via the JANUS software provided by the CAEN.

\begin{figure}
    \centering
    \includegraphics[width=1\linewidth]{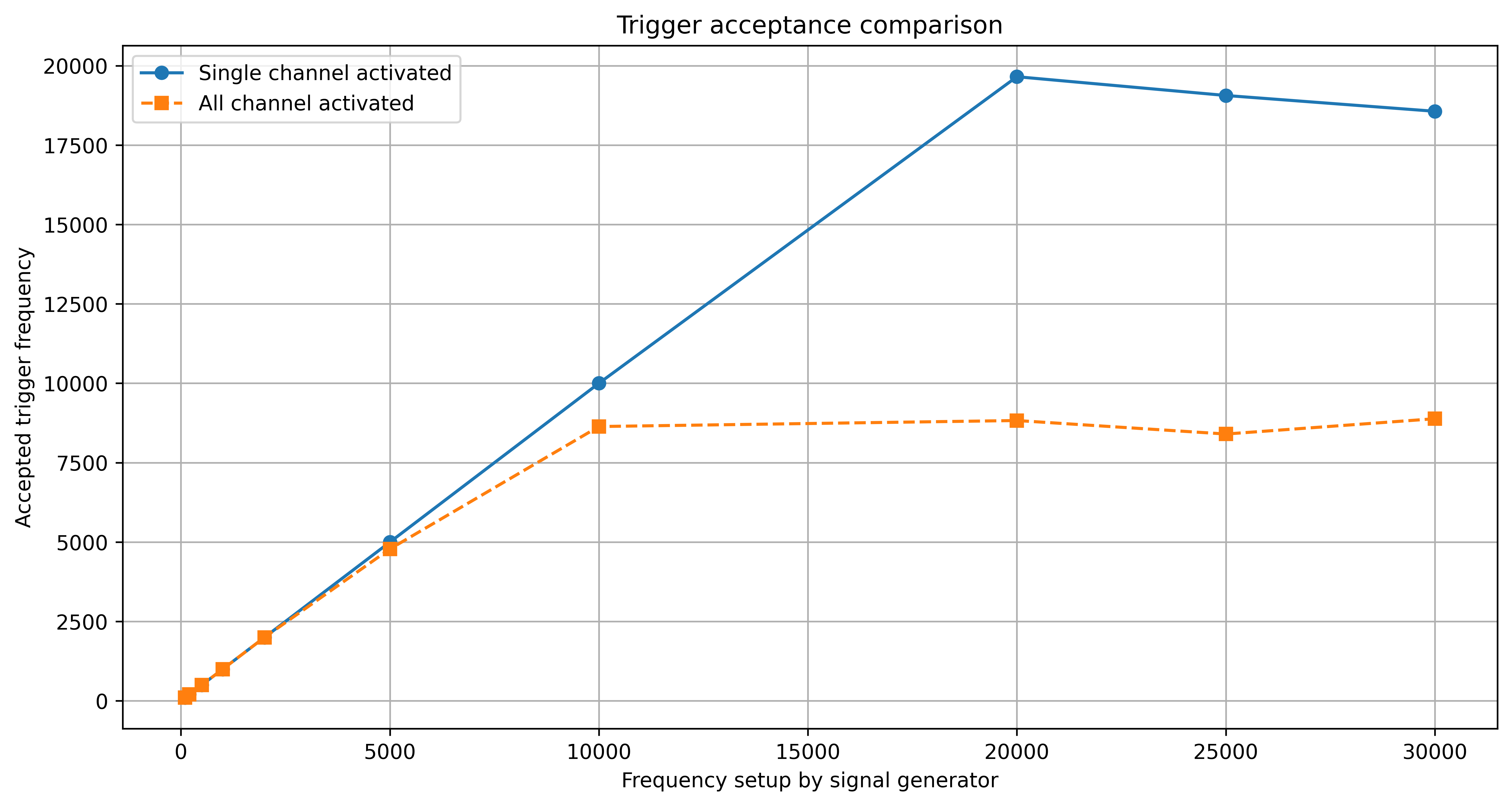}
    \caption{Trigger acceptance as a function of input frequency for two configurations: blue curve represents operation with only one active channel, while the orange curve corresponds to operation with all 64 channels active.}
    \label{fig:acceptance_freq}
\end{figure}

Figure~\ref{fig:acceptance_freq} displays the trigger acceptance performance of the FERS-5202 board under both conditions. When only a single channel was active, the board maintained full acceptance up to approximately 20 kHz. However, with all 64 channels active, the onset of trigger losses occurred at lower frequencies, indicating a reduced acceptance threshold likely due to increased processing load or bandwidth limitations.

\section{Test with SiPMs and External Trigger}

\subsection{Single Photoelectron Spectrum Characterization}

To evaluate the FERS 5202 readout performance at low signal levels, single photoelectron (1 pe) spectrum measurements were performed using a Hamamatsu MPPC S13360-3050PE silicon photomultiplier in complete darkness. The S13360-3050PE, with its 3×3 mm² active area and 50 $\mu$m  pixel pitch, provides an ideal test case for low-light detection capabilities due to its inherent dark count rate and well-defined single photoelectron response.

\begin{figure}
    \centering
    \includegraphics[width=1\linewidth]{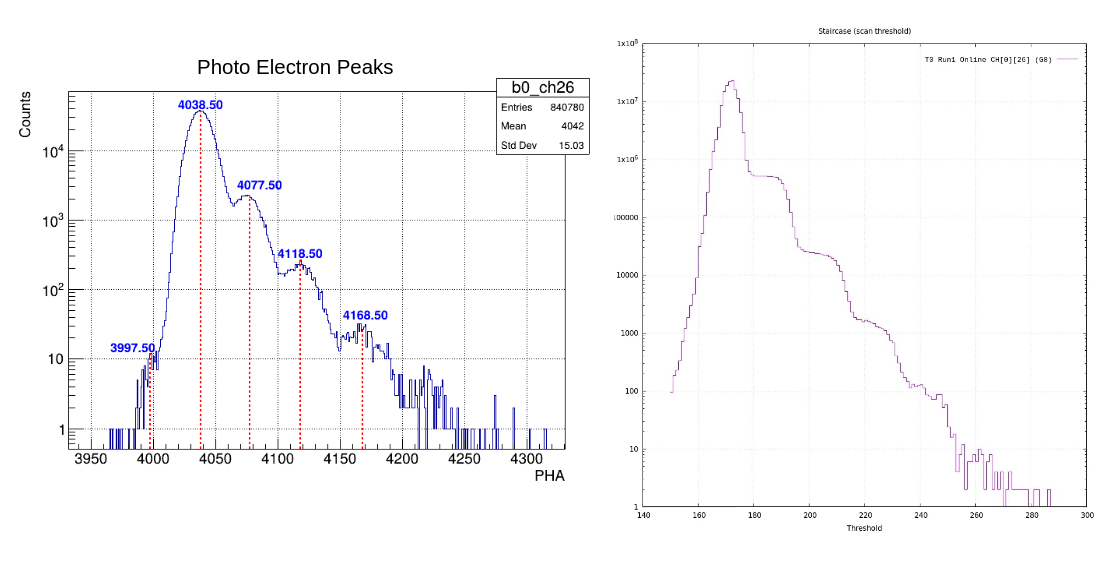}
    \caption{Photoelectron spectrum obtained from the SiPM detector. The y-axis is plotted on a logarithmic scale to highlight both low and high-count regions (left). Measured count rate versus discriminator threshold for the SiPM The stepwise decrease in counts corresponds to successive suppression of signals from dark noise and discrete photoelectron peaks (right).}
    \label{fig:triple_gaus}
\end{figure}
\begin{figure}
    \centering
    \includegraphics[width=0.8\linewidth]{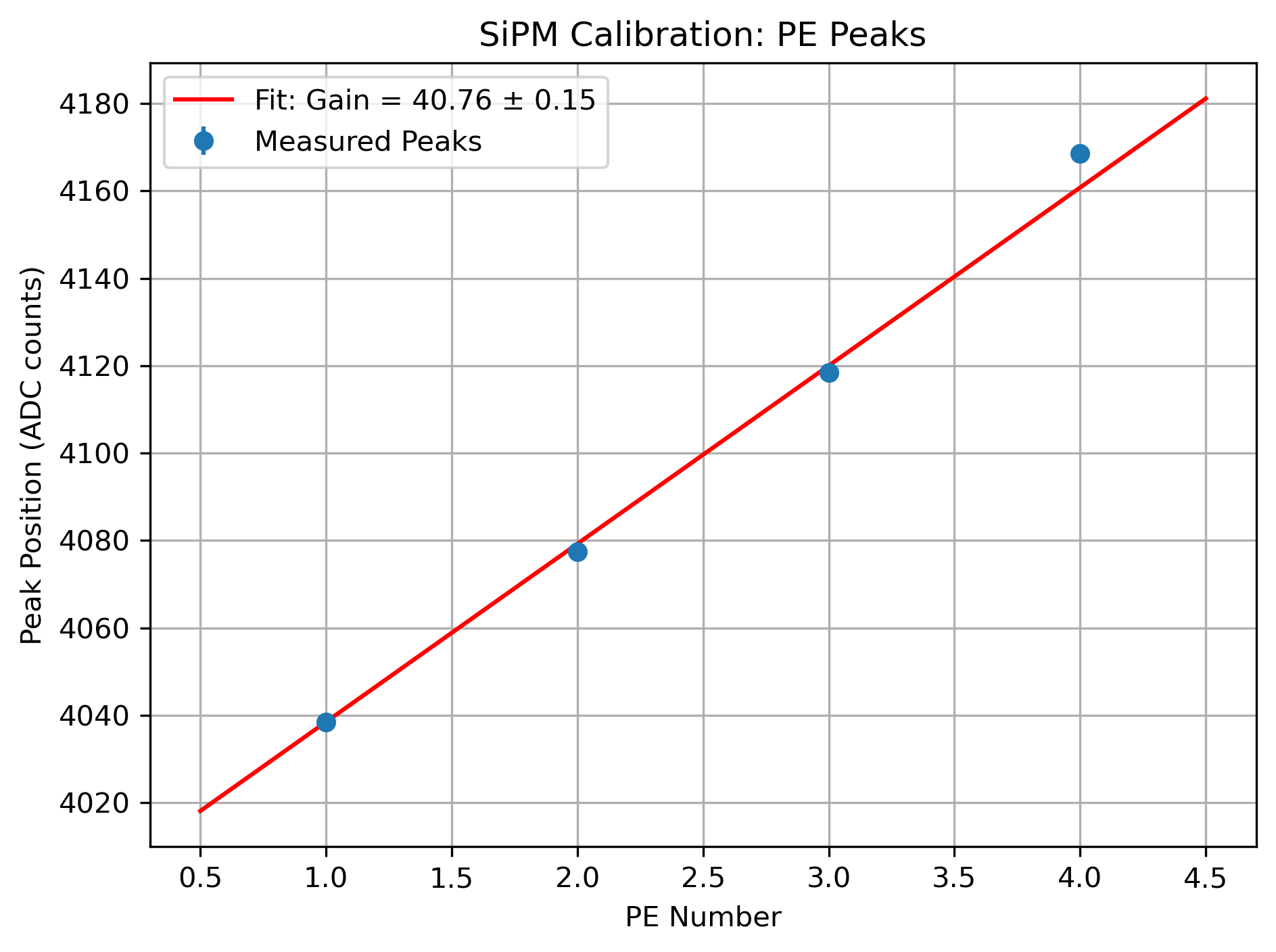}
    \caption{SiPM detector calibration through multi-photoelectron peak analysis. Peak positions for 1–4 photoelectrons (PE) are shown with statistical error bars. Linear fitting (red line) determines a single-photoelectron gain of 40.76 ± 0.15 PHA units/p.e. and establishes the pedestal baseline. The fit slope quantifies the charge-to-Peck Hight Amplitude (PHA) conversion factor per photoelectron, while the intercept corrects for the detector's baseline offset.}
    \label{fig:SiPM_Calib}
\end{figure}

  The \textit{stair case} plot (figure ~\ref{fig:triple_gaus}(right)) is obtained by measuring the number of signals above the threshold as a function of the discriminator threshold value. It is used to calibrate  in p.e. the threshold value of the discriminator, allowing the photoelectron cut-off corresponding to a certain threshold value to be established.  This is particularly useful when the FERS operates in self-triggered mode. For example the dark noise spectrum shown in the figure~\ref{fig:triple_gaus}(left) was obtained in self-triggered mode with a threshold value that removes the electronic noise of the system and allows to acquire the 1 p.e. and multi-photoelectron events.
  The high-gain parameter was set to 20 arbitrary units within the JANUS GUI. To ensure clear signal separation, the pedestal level was adjusted so that the digitised amplitude of the peak height (PHA) corresponds to 4000, in arbitrary units. The peaks obtained for 15 minutes run for SiPM 1 is shown in the figure~\ref{fig:triple_gaus}(left). The number of counts axis is in Log scale for the easy visibility. The dark spectrum was obtained for the SiPM and the values for 1st four peaks are mentioned in the table~\ref{tab:1pe_table}. The single p.e. value can be obtained by the linear fitting of the peaks as shown in the figure~\ref{fig:SiPM_Calib}.

\begin{table}[ht]
\centering
\caption{photo electron peak values }
\begin{tabular}{|c |c |c | c | c |c |}
\hline
    & pedestal (0 p.e) & 1 p.e & 2 p.e & 3 p.e & 4 p.e \\ \hline
SiPM  & $3997.50 \pm 0.06$  & $4038.50 \pm 0.04$ & $4077.50 \pm 0.57$ & $4118.50 \pm 0.32$ & $4168.50 \pm 0.88$\\ \hline

\end{tabular}

\label{tab:1pe_table}
\end{table}

\subsection{Cosmic Muon with external trigger}
A scintillation detector system was developed to characterize the performance of the FERS 5202 readout using cosmic-ray muons as a natural radiation source. The experimental setup incorporated both photomultiplier tubes (PMTs) and silicon photo-multipliers (SiPMs) to enable coincidence-based muon detection and subsequent analysis of the SiPM readout chain.

\begin{figure}
    \centering
    \includegraphics[width=0.9\linewidth]{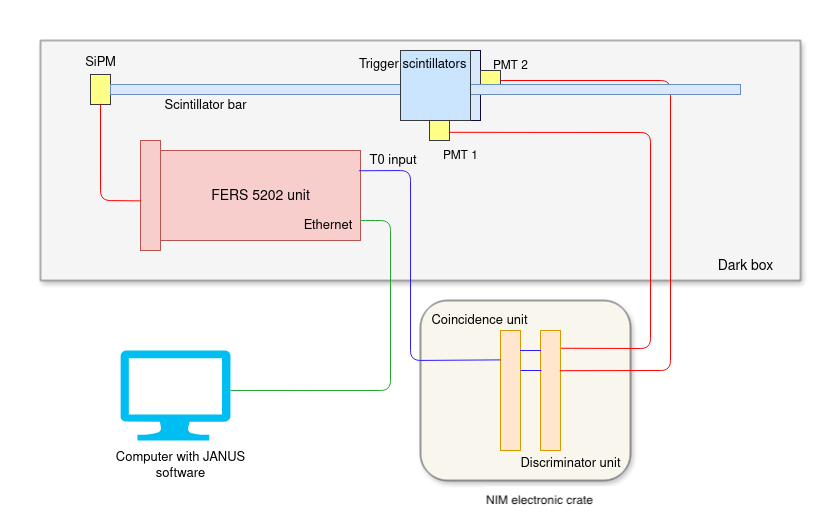}
    \caption{Sketch of the experimental setup, Note that power supply units for both A5202 unit and fast PMTs are not included in the sketch.}
    \label{fig:setup_darkbox}
\end{figure}

The trigger system consisted of two Hamamatsu R6094 PMTs, each optically coupled to a 4 × 4 × 0.8 cm³ plastic scintillator paddle positioned to define the geometric acceptance for cosmic muons. A Hamamatsu MPPC S13360-3050PE SiPM was coupled to a 1 m-long plastic scintillator bar (0.8 × 0.6 cm² cross-section) placed between the PMT trigger elements. A sketch of the experimental setup is shown in figure .~\ref{fig:setup_darkbox}.

\begin{figure}
    \centering
    \includegraphics[angle=270, width=0.7\linewidth]{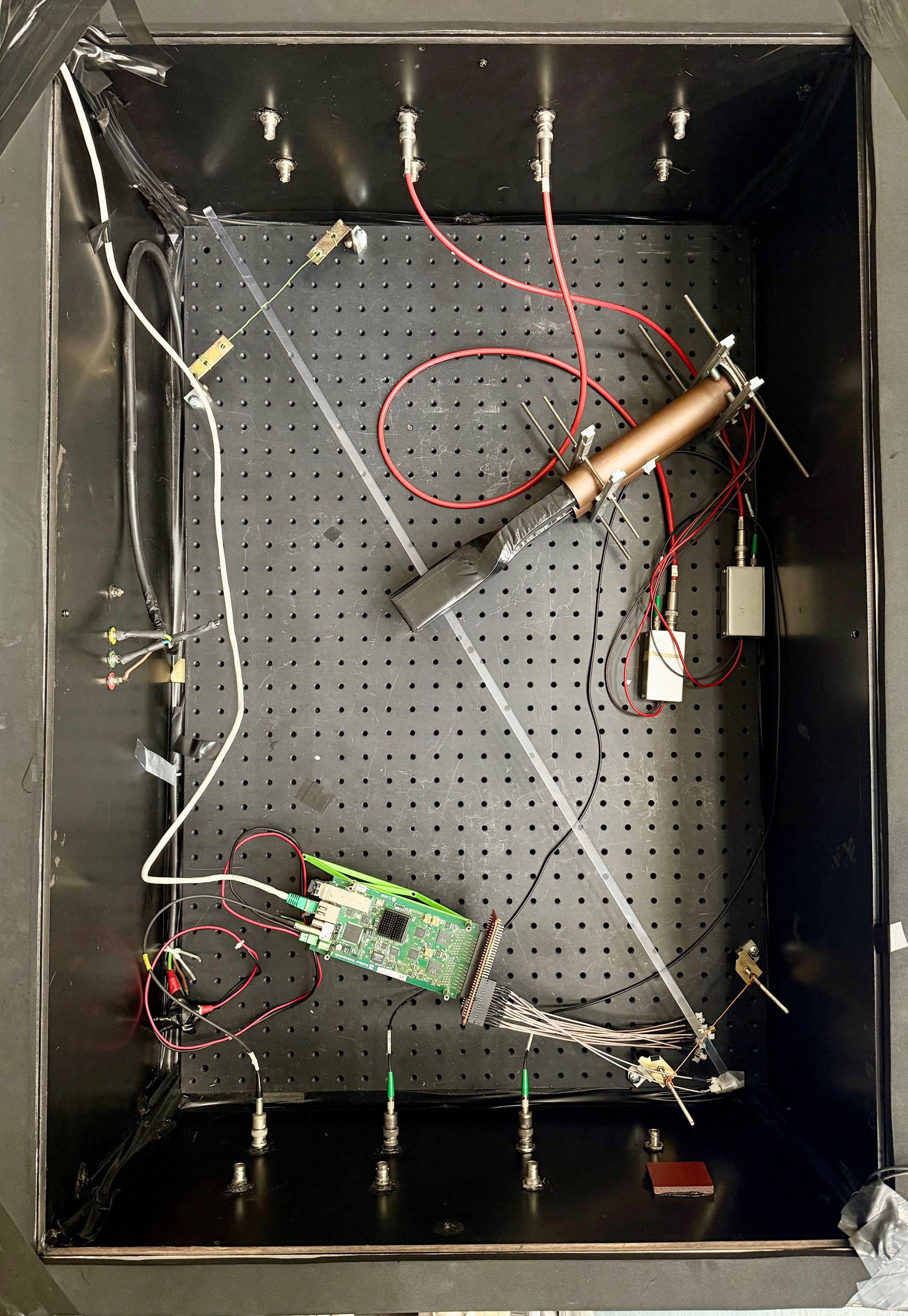}
    \caption{Experimental setup assembled inside the light-tight enclosure. Scintillator bar + SiPM are connected to the FERS board.}
    \label{fig:dark_box}
\end{figure}

To minimize optical contamination, the entire detector assembly—including both PMTs, the SiPM, and the scintillators—was housed within a light-tight enclosure (see figure.~\ref{fig:dark_box}). The PMT signals were extracted from the dark box and conditioned using a CAEN N841 16-channel discriminator, with the outputs fed into a CAEN N455 coincidence logic unit. These electronics modules were located outside the enclosure to facilitate operation and monitoring. The resulting coincidence signal provided a clean trigger for muon events, which was routed to the T0 LEMO input of the FERS 5202 readout system. This triggering scheme effectively suppressed background events while maintaining high efficiency for cosmic muon detection.

This configuration enabled time-correlated measurements between the PMT trigger system and the SiPM response, providing an ideal testbed for evaluating the performance of the FERS 5202 with realistic detector signals. Cosmic muon events provide a well-characterized energy deposition profile, allowing for a comprehensive assessment of the readout electronics’ linearity, timing resolution, and signal-processing capabilities under practical operating conditions.

The timing characteristics of the detection system required careful consideration due to the inherent differences in signal propagation times. The Hamamatsu R6094 PMTs exhibit a transit time of approximately 30 ns~\cite{Hamamatsu_R6094}, while the MPPC S13360-3050PE responds within picoseconds. Combined with the processing delays through the discriminator and coincidence modules plus cable delays, the total trigger signal delay approached 90 ns. Since the FERS 5202's programmable slow shaper can be configured up to 87.5 ns, the SiPM signal arrived at the readout board before the PMT-derived trigger signal.
To address this timing mismatch, the FERS 5202's validation trigger mode was employed through the JANUS software interface. In this configuration, the system operates with internal self-triggering when the SiPM signal exceeds the programmed threshold, while the PMT coincidence signal fed to the T0 input serves as a validation gate. This approach ensures that only SiPM signals coincident with valid muon events detected by the PMT trigger system are recorded and processed.
This hybrid triggering scheme successfully enabled the acquisition of cosmic muon energy spectra while maintaining the benefits of both the SiPM's fast response and the PMT system's reliable particle identification. The configuration provided an effective testbed for evaluating the FERS 5202's performance with realistic detector signals and timing requirements.

\begin{figure}
    \centering
    \includegraphics[width=0.95\linewidth]{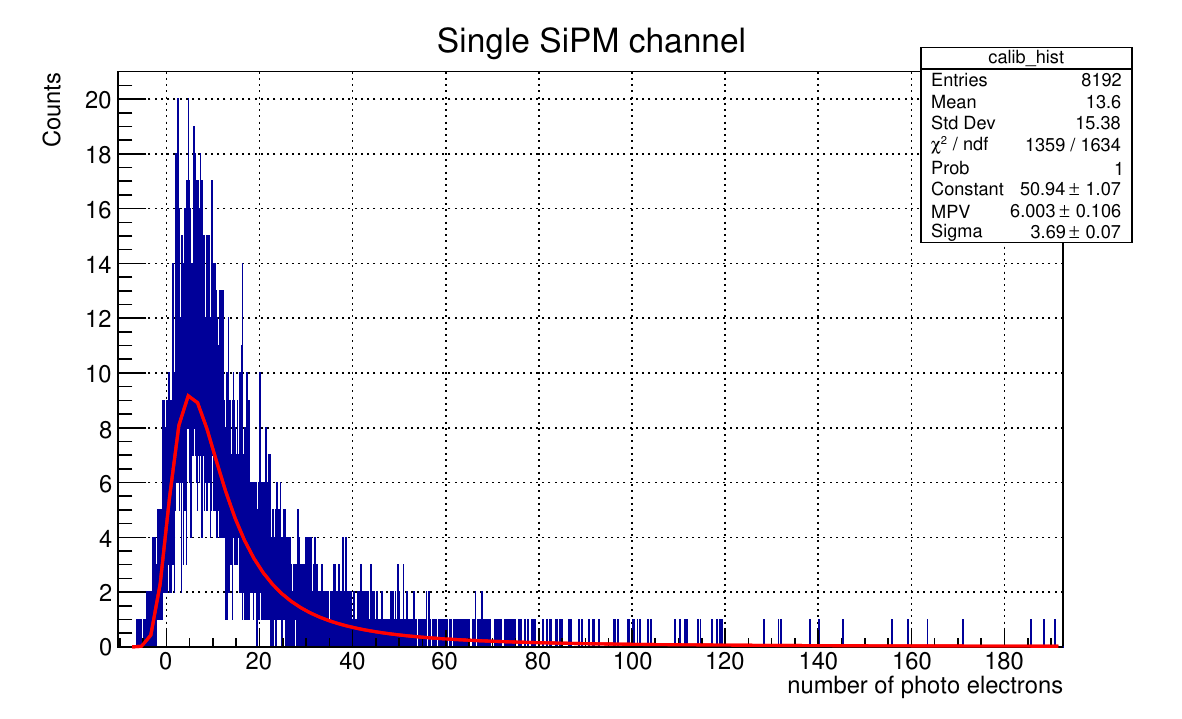}
    \caption{Muon energy deposition measured in spectroscopy mode. The red curve shows a Landau distribution fitted to the photoelectron yield detected by the SiPM.}
    \label{fig:Muon_energy}
\end{figure}

This hybrid triggering scheme successfully enabled the acquisition of cosmic muon energy spectra while maintaining the benefits of both the SiPM's fast response and the PMT system's reliable particle identification. The resulting muon energy spectrum from the SiPM-scintillator bar system is presented in Figure~\ref{fig:Muon_energy}. Due to the geometric constraints and limited acceptance area of the PMT trigger system, continuous data acquisition over four days was required to accumulate sufficient statistics for the energy spectrum analysis. This extended measurement period demonstrates the system's stability and the FERS 5202's capability for long-term data acquisition under realistic experimental conditions.

\section{Conclusion}

This comprehensive characterization study of the FERS 5202 front-end readout unit demonstrates its suitability for a wide range of silicon photomultiplier applications, from single photoelectron detection to cosmic-ray particle measurements. The evaluation encompassed both fundamental performance metrics and practical implementation scenarios using Hamamatsu MPPC devices.

The single photoelectron spectrum measurements using the S13360-3050PE MPPC in dark conditions established the readout system's capability to resolve minimal charge signals corresponding to individual photoelectron avalanches. The clear separation of the pedestal peak from the single photoelectron peak demonstrates the FERS 5202's excellent charge resolution and low electronic noise performance at the fundamental detection limit.

The cosmic muon detection experiment provided a rigorous test of the system's performance under realistic experimental conditions. The successful implementation of a hybrid PMT-SiPM trigger system, despite the timing requirements, showcases the flexibility of the FERS 5202's trigger validation capabilities. The four-day continuous data acquisition period required to accumulate sufficient muon statistics not only demonstrates the system's long-term stability but also validates its reliability for extended experimental campaigns.

The ability to handle the timing mismatch between fast SiPM signals and slower PMT-derived triggers through the validation trigger mode highlights the sophisticated trigger logic capabilities of the FERS 5202. This functionality proves essential for multi-detector coincidence experiments where different detector technologies with varying response characteristics must be integrated.

Overall, the FERS 5202 front-end unit exhibits robust performance characteristics that make it well-suited for SiPM-based detection systems across diverse applications. Its combination of low-noise single photoelectron sensitivity, flexible trigger logic, and proven long-term operational stability positions it as a valuable tool for experimental physics, medical imaging, and other photon detection applications requiring precise charge measurement and timing capabilities.

The small size and high density of the channels allow the FERS to be positioned close to the detector. 
This minimizes noise pickup and signal attenuation are critical factors for bore hole deployments where signal integrity must be maintained in electrically noisy environments and over the extended cable lengths often required for deep geological investigations.

\acknowledgments
We would like to thank CAEN for providing us with two FERS cards for viewing and Dr Vincenzo Bottiglieri for his suggestions on preparing the experimental setup.


\newpage
\bibliographystyle{JHEP}
\bibliography{biblio.bib}

\end{document}